\documentclass[12pt]{article}
\usepackage{microtype}

\usepackage{geometry}
 \usepackage[usenames]{color}

\usepackage{authblk}
 
\usepackage{amsthm,amsfonts,amssymb,amsmath,amscd,bezier}
\usepackage{graphicx,hyperref, latexsym,wrapfig,makeidx}
\usepackage[all]{xy}

\usepackage{mathtools}

\usepackage{cmbright}
\usepackage[OT1]{fontenc}
\usepackage{tgbonum}
\usepackage{dsfont}
\usepackage{soul}

\usepackage{tikz}
\usepackage{tikz,tikz-qtree,tikz-qtree-compat,fullpage,adjustbox}

\usepackage{pst-plot}
\usepackage{etoolbox}

\theoremstyle{definition}
\newtheorem{theorem}{Theorem}[section]
\newtheorem{corollary}[theorem]{Corollary}
\newtheorem{conjecture}[theorem]{Conjecture}
\newtheorem{lemma}[theorem]{Lemma}
\newtheorem{proposition}[theorem]{Proposition}
\newtheorem{definition}[theorem]{Definition}
\newtheorem{remark}[theorem]{Remark}
\newtheorem{example}[theorem]{Example}

\providecommand{\keywords}[1]
{
  \small	
  \textbf{\textit{Keywords---}} #1
}

\definecolor{lime}{HTML}{A6CE39}
\DeclareRobustCommand{\orcidicon}{%
	\begin{tikzpicture}
	\draw[lime, fill=lime] (0,0) 
	circle [radius=0.16] 
	node[white] {{\fontfamily{qag}\selectfont \tiny ID}};
	\draw[white, fill=white] (-0.0625,0.095) 
	circle [radius=0.007];
	\end{tikzpicture}
	\hspace{-2mm}
}

\foreach \x in {A, ..., Z}{%
	\expandafter\xdef\csname orcid\x\endcsname{\noexpand\href{https://orcid.org/\csname orcidauthor\x\endcsname}{\noexpand\orcidicon}}
}

\title{Infinite chains in the tree of numerical semigroups}
\author{Mariana Rosas-Ribeiro%
  \thanks{Electronic address: \texttt{mariana.rosas@urv.cat}; Corresponding author} \orcidA{}}
\affil{Department of Computer Engineering and Mathematics,\\ Universitat Rovira i Virgili,\\Tarragona, Spain}

\author{Maria Bras-Amorós%
  \thanks{Electronic address: \texttt{maria.bras@upc.edu}} \orcidB{}}
\affil{Department of Mathematics,\\ Universitat Politècnica de Catalunya,\\Barcelona, Spain}

\date{January 29, 2024}
\begin{document}

\maketitle

\begin{abstract}
One major problem in the study of numerical semigroups is determining the growth of the semigroup tree. In the present work, infinite chains of numerical semigroups in the semigroup tree, firstly introduced by Bras-Amorós and Bulygin (Semigroup Forum, 71:561--574, 2009), are studied. Computational results show that these chains are rare, but without them the tree would not be infinite. It is proved that for each genus $g\geq 5$ there are more semigroups of that genus not belonging to infinite chains than semigroups belonging. Bras-Amorós and Bulygin (Semigroup Forum, 71:561--574, 2009) presented a characterization of the semigroups that belong to infinite chains in terms of the coprimality of the left elements of the semigroup as well as a result on the cardinality of the set of infinite chains to which a numerical semigroup belongs in terms of the primality of the greatest common divisor of these left elements. We revisit these results and fix an imprecision on the cardinality of the set of infinite chains to which a semigroup belongs in the case when the greatest common divisor of the left elements is a prime number. We then look at infinite chains in subtrees with fixed multiplicity. When the multiplicity is a prime number there is only one infinite chain in the tree of semigroups with such multiplicity. When the multiplicity is $4$ or $6$ we prove a self-replication behavior in the subtree and prove a formula for the number of semigroups in infinite chains of a given genus and multiplicity $4$ and $6$, respectively.
\end{abstract}

   
\section{Introduction}

A numerical semigroup $\Lambda$ is a subset of the non-negative integers $\mathbb{N}_0$, containing $0$, closed under addition and with finite complement in $ \mathbb{N}_0$. The elements of $ \mathbb{N}_0$ that are not in $\Lambda$ are called the \textit{gaps} and the number of gaps is the \textit{genus} $g(\Lambda)$ of the numerical semigroup. Intuitively, the possibilities of obtaining numerical semigroups with $g$ gaps seem to increase as the number $g$ increases and indeed computational results have shown this (\cite{m-fibonacci}, \cite{fromentin-tree}). It was conjectured in \cite{m-fibonacci}, 2008, that the sequence $n_g$ formed by the number of numerical semigroups of each genus $g$, not only is increasing but also grows like the Fibonacci sequence. In 2012, one such conjecture was proved \cite{zhai}, namely $\dfrac{n_g}{\varphi^g}\xrightarrow{g \to \infty }S$, where $\varphi$ is the golden ratio and $S$ is at least $3.78$. It remains unproved that $n_{g-2}+n_{g-1}\leq n_g$; even the weaker conjecture that $n_{g-1}\leq n_g$, already announced in 2007 \cite{m-segovia}, remains open.

One way to label all elements of the numerical semigroup $\Lambda$ is to enumerate them. The enumeration $\lambda$ of a numerical semigroup $\Lambda$ is the unique increasing bijection between $ \mathbb{N}_0$ and $\Lambda$, and we denote by $\lambda_i$ the image of $i\in \mathbb{N}_0$. In particular, $\lambda_0=0$. The \textit{multiplicity} of $\Lambda$ is $m(\Lambda)=\lambda_1$. It holds $m(\Lambda)\leq g(\Lambda)+1$. \textit{The \textit{conductor} of $\Lambda$} is $c(\Lambda)=\lambda_{k}$, where $\lambda_{k}+n\in \Lambda$ for all $n\in \mathbb{N}$ and $\lambda_{k}-1 \notin \Lambda$. The number $c(\Lambda)-1$ is the largest gap and is called the \textit{Frobenius number} of $\Lambda$ and denoted by $F(\Lambda)$. The elements of $\Lambda$ that are smaller than its conductor are the \textit{left elements}, denoted by $ \mathcal{L}(\Lambda)$.

Let $[a,\infty)$ stand for $\{i\in {  \mathbb N}_0: i\geq a\}$. Numerical semigroups of the form $\{0\}\cup[\lambda_1,\infty)$ are called \textit{ordinary} and those of the form $\{2n:n\geq0\}\cup[\lambda_{k},\infty), \ \lambda_{k}$ even, are called \textit{hyperelliptic semigroups}. 
 
It is well known that each numerical semigroup has a unique minimal system of generators. The cardinality of the minimal system of generators is called the \textit{embedding  dimension} of $\Lambda$ and we denote it by $e(\Lambda)$. In view of the fact that two generators cannot lie in the same congruence class modulo $m(\Lambda)$, one has $e(\Lambda)\leq \lambda_1$. The minimal generators of $\Lambda$ that are not in $  \mathcal{L}(\Lambda)$ are the \textit{effective generators} or \textit{right generators} of $\Lambda$ and with them we can organize the collection of all numerical semigroups in a tree rooted at $ \mathbb{N}_0$. In Figure~\ref{fig:tree} one can see the tree with the nodes of level at most 6. 
      \begin{figure}\input{inputfile-list-semigrouptree-6.tex}
        \caption{Tree of numerical semigroups with the nodes of level at most 6.}
        \label{fig:tree}
        \end{figure}
 In this representation, the elements in orange are the effective generators. That tree is constructed by removing one by one the effective generators from each numerical semigroup. In this way, the level of the tree is equal to the genus of the numerical semigroups that are at that level (assume that the root has level zero). For more details of this tree, we suggest consulting \cite{m-bounds}, \cite{zhai} and \cite{fromentin-tree}. From the tree perspective, numerical semigroups can also be seen as nodes. In this way the \textit{children} of a numerical semigroup $\Lambda$ are those numerical semigroups that are obtained by taking away one right generator from $\Lambda$. Conversely, the \textit{parent} of a numerical semigroup is obtained by adding its Frobenius number (the parent is one level lower in the tree). The number of children of a node $\Lambda$ is called its \textit{efficacy} and is denoted by $h(\Lambda)$. By the previous remarks, one can deduce that $h(\Lambda)\leq e(\Lambda) \leq m(\Lambda)\leq g(\Lambda)+1$. As can be seen in the references cited above, this tree has been used to study the growth of the $n_g$ sequence.

 Every numerical semigroup of genus larger than $g(\Lambda)$ connected to $\Lambda$ through edges in the tree is said to be a \textit{descendant} of $\Lambda$. A node that has no descendants is said to be a \textit{leaf} and a node that has exactly one child is said to be a \textit{stick}. An infinite set of numerical semigroups is said to be an \textit{infinite chain} if it contains ${\mathbb N}_0$ and the semigroups in the set can be enumerated in such a way that each one is the parent of the next one.
 
The set of ordinary semigroups and the set of hyperelliptic semigroups each form an infinite chain and, as we will see throughout this paper, these two chains are well positioned in the numerical semigroup tree.
 
Infinite chains of numerical semigroups were first investigated in \cite{m-towards}. This reference presented a characterization of the semigroups that belong to infinite chains in terms of the coprimality of the left elements of the semigroup as well as a result on the number of infinite chains to which a numerical semigroup belongs in terms of the primality of the greatest common divisor of these left elements.

In Section~\ref{section:preliminary} we state some preliminary results and in Section \ref{section-infinite-chains} we revisit the main result of \cite{m-towards} and fix an imprecision on the cardinality of the set of infinite chains to which a semigroup belongs in the case when the greatest common divisor of the left elements is a prime number. The resulting fixed theorem is Theorem~\ref{theorem:infinitechains}.

Computational results show that infinite chains are rare, but without them the tree would not be infinite. For this reason, we direct the study on infinite chains in order to investigate the broader problem of the tree growth. In Section~\ref{section:majoritynoninfinitechains}
we prove that for each genus $g\geq 5$ there are more semigroups of that genus not belonging to infinite chains than semigroups belonging (Theorem~\ref{theorem:majority}).

Another aspect that is evident in the tree is that semigroups of the same multiplicity form subtrees. In Section~\ref{sec:multi} we look at infinite chains in subtrees with fixed multiplicity. When the multiplicity is a prime number, one can prove, as a consequence of Theorem~\ref{theorem:infinitechains}, that there is only one infinite chain in the tree of semigroups with such multiplicity. For the cases of multiplicity $4$ and multiplicity $6$, we prove a self-replication behavior in the subtree formed by the semigroups with fixed multiplicity equal to $4$ and that are in infinite chains (Subsection~\ref{subsec-m4}); and a different self-replication behavior in the subtree formed by the semigroups with fixed multiplicity equal to $6$ and that are in infinite chains (Subsection~\ref{subsec-m6}). This enables us to prove a formula for the amount of semigroups of genus $g$ that belong to infinite chains, of any given prime multiplicity (Theorem~\ref{theorem:primem}), of multiplicity $4$ (Theorem~\ref{theorem:mfour}), or of multiplicity $6$ (Theorem~\ref{theorem:msix}).

\section{Preliminary results}\label{section:preliminary}

The next lemma was proved in \cite{m-bounds} but we enunciate it here in a different way, underlying the result used in the original proof. Among other things, it allows to study the children of the children, or grandchildren, of most numerical semigroups.

\begin{lemma}[\cite{m-towards}, Lemma 1]\label{lemma:effective-generators} 
If $\lambda_{i_1}<\lambda_{i_2}<\dots<\lambda_{i_n}$ are the effective generators of a non-ordinary numerical semigroup $\Lambda$, then the effective generators of $\Lambda\setminus\{\lambda_{i_j}\}$ are either $\lambda_{i_{j+1}},\dots,\lambda_{i_n}$ or $\lambda_{i_{j+1}},\dots,\lambda_{i_n},\lambda_{i_j}+\lambda_1$.
\end{lemma}

Our next proposition gives a bound for the maximum effective generator.

\begin{proposition}\label{proposition:most-effective-generator}
Every effective generator is at most the conductor plus the multiplicity minus one.
\end{proposition}

\begin{proof}
Indeed, for a numerical semigroup $\Lambda$, if $a\geq c(\Lambda)+m(\Lambda)$, then $a=m(\Lambda)+(a-m(\Lambda))$, with both $m(\Lambda)$ and $a-m(\Lambda)$ nongaps.
\end{proof}

The last preliminary result is a characterization of ordinary semigroups by the number of children in terms of the genus.

\begin{lemma}\label{lemma:ordinary-g+1-children}
If a numerical semigroup $\Lambda$ with genus $g$ has $g+1$ children, then $\Lambda$ is ordinary.
\end{lemma}

\begin{proof}
The following sequence of inequalities must indeed be a sequence of equalities: $$g+1 = h(\Lambda) \leq e(\Lambda) \leq m(\Lambda) \leq g+1.$$ Hence, $m(\Lambda)=g+1$, meaning that all gaps of $\Lambda$ are smaller than $m(\Lambda)$, and, hence, $\Lambda$ must be ordinary.
\end{proof}

\section{Infinite chains}\label{section-infinite-chains}

In this section we are interested in studying the nodes of the tree of numerical semigroups that have infinitely many descendants. An infinite chain is a sequence of semigroups $\Lambda_0,\Lambda_1,\Lambda_2,\dots$ such that $\Lambda_0=\mathbb{N}_0$ and $\Lambda_i$ is the parent of $\Lambda_{i+1}$ in the semigroup tree. Notice that a numerical semigroup belongs to an infinite chain if and only if it has infinitely many descendants. 

First, we analyze which nodes in the semigroup tree have an infinite number of descendants. For the nodes having a finite number of descendants we give a way to determine the descendant at largest distance; for the nodes having an infinite number of descendants we determine the number of infinite chains in which the semigroup lies. It turns out here that primality and coprimality of integers appear in the scene as discriminating factors. 

Let ${\mathbb S}$ be the set of all numerical semigroups, and let ${\mathbb I}$ be the set of all infinite chains. One element of ${\mathbb I}$ contains infinitely many elements in ${\mathbb S}$. As an example, the sequence of ordinary semigroups, ${\mathcal O}_g=\{0,g+1,\rightarrow\}$ for any $g\in{\mathbb N}_0$, is in ${\mathbb I}$. We denote it $I_{\mathcal O}$. That is, $${I_{\mathcal O}}=\{{\mathcal O}_g: g\geq 0\}.$$ Another example is the sequence of hyperelliptic semigroups, ${\mathcal H}_g=\{0,2,4,\dots,2g,\rightarrow\}$. We denote it $I_{\mathcal H}$. That is, $${I_{\mathcal H}}=\{{\mathcal H}_g: g\geq 0\}.$$

Each element in ${\mathbb S}$ may be contained in none, one, several, or infinitely many elements of ${\mathbb I}$. The aim of this section is  analyzing what circumstances give what of these cases. Then, we analyze what kind of generators appear in infinite chains.

For the proof of the next lemma we use that the integers $\ell_1,\dots,\ell_m$ generate a numerical semigroup if and only if they are coprime. 

\begin{lemma}
\label{lem:infinitechaintoseminteger}
Given an infinite chain $I=(\Lambda_i)_{i\geq 0}$ different than $I_{\mathcal O}$, it holds that $$\bigcap_{i\geq 0}\Lambda_i=d\cdot \Lambda$$ for some integer $d>1$ and some numerical semigroup $\Lambda$.
\end{lemma}

\begin{proof}
The intersection $\bigcap_{i\geq 0}\Lambda_i$ satisfies $0\in \bigcap_{i\geq 0}\Lambda_i$. Since there exists a non-ordinary semigroup in $I$, there must be a non-zero element $x\in \bigcap_{i\geq 0}\Lambda_i$ and $x+y\in \bigcap_{i\geq 0}\Lambda_i$ for all $x,y\in \bigcap_{i\geq 0}\Lambda_i$.

Furthermore, all elements in $\bigcap_{i\geq 0}\Lambda_i$ must be divisible by an integer $d>1$. Indeed, otherwise we could find a finite set of coprime elements which would generate a numerical semigroup, and this numerical semigroup should be a subset of $\bigcap_{i\geq 0}\Lambda_i$. Then the infinite chain would not contain any semigroup with genus larger than that of this semigroup, giving a contradiction. Let $d$ be the greatest of the common divisors of $\bigcap_{i\geq 0}\Lambda_i$. Then $\frac{1}{d}\left(\bigcap_{i\geq 0}\Lambda_i\right)$ must be a numerical semigroup.
\end{proof}

\begin{lemma}
\label{lem:semintegertoinfinitechain}
Given an integer $d>1$ and a numerical semigroup $\Lambda$, the infinite chain obtained by deleting repetitions in the sequence $\Lambda_j=d\cdot \Lambda\cup\{\ell\in{\mathbb N}:\ell\geq j\}$ has intersection $d\cdot \Lambda$.
\end{lemma}

Lemma~\ref{lem:infinitechaintoseminteger} 
and its proof suggest the map $$\begin{array}{rcl}
\sigma:{\mathbb I}\setminus I_{\mathcal O}&\longrightarrow&{\mathbb N}_{\geq 2}\times{\mathbb S}
\\
\{\Lambda_i\}_{i\in{\mathbb N}_0}&\mapsto&(\gcd(\bigcap_{i\geq 0}\Lambda_i)),
\bigcap_{i\geq 0}\Lambda_i/\gcd(\bigcap_{i\geq 0}\Lambda_i)).\\
\end{array}
$$

Lemma~\ref{lem:semintegertoinfinitechain}
proves that the map $$\begin{array}{rcl}
\omega:{\mathbb N}_{\geq 2}\times{\mathbb S}&\longrightarrow&{\mathbb I}\setminus I_{\mathcal O}
\\
(d,\Lambda)&\mapsto&
\{d\cdot \Lambda\cup\{\ell\in{\mathbb N}:\ell\geq i\}\}_{i\not\in d\cdot\Lambda}\end{array}
$$ 
is the inverse of $\sigma$.

Consequently, ${\mathbb I}\setminus I_{\mathcal O}$
and ${\mathbb N}_{\ge 2}\times{\mathbb S}$ are in a one-to-one correspondence. For example, in this correspondence, the image of $I_{\mathcal H}$ would be $(2,{\mathbb N}_0)$.

\begin{remark}
Any ordinary numerical semigroup belongs to infinitely many infinite chains. Indeed, $\{0,m,\rightarrow\}$ belongs, among others to $\omega(m',{\mathbb N}_0)$ for any $m'\geq m$.
\end{remark}

In the next theorem we show that the greatest common divisor of the left elements of a non-ordinary numerical semigroup determines whether the numerical semigroup has an infinite number of descendants. We say that a descendant of a numerical semigroup is a descendant {\em beyond} a given element $a$ of its parent if it contains all semigroup elements of the parent up to $a$. For example, $\{0,4,5,8,9,10,12,\rightarrow\}$ and $\{0,4,5,6,8,\rightarrow\}$ are descendants of $\{0,4,\rightarrow\}$, but only $\{0,4,5,6,8,\rightarrow\}$ is a descendant beyond $6$.

We use the fact that a finite number of coprime elements generate a numerical semigroup.

\begin{theorem}
\label{theorem:infinitechains}
Let $\Lambda$ be a non-ordinary numerical semigroup with enumeration $\lambda$, genus $g$, and conductor $c$, and let $d$ be the greatest common divisor of $\mathcal{L}(\Lambda)$. Then,
  \begin{enumerate}
  \item 
    $\Lambda$ lies in an infinite chain if and only if $d\neq 1$.
  \item 
    If $d=1$, then the descendant of $\Lambda$ with largest genus
    is the numerical semigroup generated by  the non-zero elements of $\mathcal{L}(\Lambda)$, that is, the numerical semigroup generated by $\lambda_1,\dots,\lambda_{c-g-1}$. 
  \item 
    If $d\neq 1$ and $d$ is not prime, then 
    $\Lambda$ lies in infinitely many infinite chains.
  \item If $d$ is a prime, then the number of infinite chains in which $\Lambda$ lies is one plus the number of descendants of
    $\Lambda/d:=\{0,\frac{\lambda_{1}}{d},\dots,\frac{\lambda_{c-g-1}}{d}\}\cup \{\ell\in{\mathbb N}_0: \ell \geq \lceil\frac{c}{d}\rceil\}$
    beyond $\frac{\lambda_{c-g-1}}{d}$.
  \end{enumerate}
\end{theorem} 

\begin{proof}

  \begin{enumerate}
  \item
  If $c-g-1=1$, then $\Lambda$ belongs at least to the infinite chain $\omega(\lambda_1,{\mathbb N}_0)$, while $d=\lambda_1\neq 1$. So, we can assume that $c-g-1\geq 2$.
      
If $d=1$ and $c-g-1\geq 2$, then $\lambda_1,\dots,\lambda_{c-g-1}$ generate a numerical semigroup $\Lambda'$ with no descendants in the semigroup tree.
Now, each descendant of $\Lambda$ must contain $\Lambda'$. Thus, the maximum of the genera of the descendants of $\Lambda$ is the genus of $\Lambda'$, which is finite, and so there is no infinite chain containing $\Lambda$.

On the other hand, if $d\neq 1$ and $c-g-1\geq 2$, then
$$\lambda_0=d\tilde\lambda_0,\lambda_1=d\tilde\lambda_1,\dots,\lambda_{c-g-1}=d\tilde\lambda_{c-g-1}$$ with 
$\tilde\lambda_1,\dots,\tilde\lambda_{c-g-1}$ coprime.
Let $\tilde\Lambda$ be the numerical semigroup generated by $\tilde\lambda_1,\dots,\tilde\lambda_{c-g-1}$. Then, $\Lambda$ belongs to the infinite chain $\omega(d,\tilde\Lambda)$.
\item As we have seen in the previous statement,
the semigroup $\Lambda'$, which is a descendant itself of $\Lambda$, must be contained in each descendant of $\Lambda$. 
\item If $d$ is not prime, then $d=d_1d_2$ for some $d_1,d_2>1$ and, as before, 
$$
  \lambda_0=d_1d_2\tilde\lambda_0,\lambda_1=d_1d_2\tilde\lambda_1,  \dots,\lambda_{c-g-1}=d_1d_2\tilde\lambda_{c-g-1}
$$  with $\tilde\lambda_1,\dots,\tilde\lambda_{c-g-1}$ coprime. Let $\tilde\Lambda$ be ${\mathbb N}_0$ if $c-g-1=1$ or the numerical semigroup generated by
$\tilde\lambda_1,\dots,\tilde\lambda_{c-g-1}$, otherwise. For each $a\geq 0$
define $$\Lambda_{a}=d_2\tilde\Lambda\cup \{\ell\in{\mathbb N}_0:\ell\geq a\}.$$ Then $\Lambda$ belongs to the infinite chains $\omega(d_1,\Lambda_a)$ for all $a>\frac{c}{d_1}$ that are not multiples of $d_2$, which are all different. So, it belongs to an infinite number of infinite chains.
\item Suppose that $d$ is prime and that an infinite chain $(\Lambda_i)_{i\geq 0}$ contains $\Lambda$. The infinite chain must satisfy $\bigcap_{i\geq 0}\Lambda_i=d\cdot\tilde{\Lambda}$ for a numerical semigroup 
$\tilde{\Lambda}$ such that
\begin{itemize}
\item 
  $d\tilde{\lambda}_0=\lambda_0,d\tilde{\lambda}_1=\lambda_1, \dots d\tilde{\lambda}_{c-g-1}={\lambda}_{c-g-1}$,
\item
$d\tilde{\lambda}_{c-g}\geq c$, since $d\tilde{\Lambda}\subseteq \Lambda$.
\end{itemize}
Now, the conductor $\tilde c$ of $\tilde{\Lambda}$ must be $\tilde c\geq \tilde\lambda_{c-g}\geq \lceil\frac{c}{d}\rceil$. Thus, $\tilde{\Lambda}$ is either 
$\Lambda/d$
or one of its descendants beyond $\frac{\lambda_{c-g-1}}{d}$.

Conversely, if $\tilde\Lambda$ is either $\Lambda/d$ or one of its descendants beyond $\frac{\lambda_{c-g-1}}{d}$, then $\Lambda$ belongs to the infinite chain $\omega(d,\tilde\Lambda)$.

It remains to prove that $\Lambda/d$ has a finite number of descendants beyond $\frac{\lambda_{c-g-1}}{d}$. If $c-g-1=1$, then $d=\lambda_1\neq 1$ and $\frac{\lambda_{c-g-1}}{d}=1$, and the semigroup
$\Lambda/d$ is ${\mathbb N}_0$, which has no descendants beyond $1$. On the other hand, if $c-g-1\geq 2$, then any descendant beyond $\frac{\lambda_{c-g-1}}{d}$ contains $\frac{\lambda_{1}}{d},\dots,\frac{\lambda_{c-g-1}}{d}$ which is a set of at least two elements which are coprime. Hence, any such descendant contains the numerical semigroup generated by $\frac{\lambda_{1}}{d},\dots,\frac{\lambda_{c-g-1}}{d}$. Now, the result follows from the fact that the number of semigroups that contain a given numerical semigroup is finite.
  \end{enumerate}
\end{proof}

\section{Minority of semigroups in infinite chains}\label{section:majoritynoninfinitechains}

In this section we prove that for each fixed genus $g\geq 5$ the majority of semigroups do not belong to any infinite chain.

From the characterization given by Theorem~\ref{theorem:infinitechains} of the semigroups that are in infinite chains, we are able to investigate which children of such semigroups remain in infinite chains.

\begin{corollary}\label{corollary:infinite-chain}
If the numerical semigroup $\Lambda$ lies in an infinite chain, then it has at most two children in infinite chains. 

\end{corollary}

\begin{proof}
Let $\Lambda$ be a numerical semigroup belonging to an infinite chain, $c$ its conductor, $  \mathcal{L}(\Lambda)$ its set of left elements and $d=\text{gcd}(  \mathcal{L}(\Lambda))$. By Theorem~\ref{theorem:infinitechains}, $d\neq 1$ and by definition an effective generator of $\Lambda$ is $c+i$ for some $i\geq0$. So every child of $\Lambda$ is of the type $\Lambda\setminus\{c+i\}$ for some $i\geq 0$ and the left elements of this child are either $  \mathcal{L}(\Lambda\setminus\{c\})=  \mathcal{L}(\Lambda)$, if $i=0$, or $  \mathcal{L}(\Lambda\setminus\{c+i\})=  \mathcal{L}(\Lambda)\cup \{c,\dots,c+i-1\}$, if $i>0$. Now, suppose $i\geq 2$. This implies $  \mathcal{L}(\Lambda\setminus\{c+i\})\supseteq\{c,c+1\}$ and, since gcd$(c,c+1)=1$, we have gcd$(  \mathcal{L}(\Lambda\setminus\{c+i\}))=1$. So, $\Lambda\setminus\{c+i\}$, for $i\geq2$, does not have infinitely many descendants.
\end{proof}

\begin{remark}\label{remark:descendants-type-c}
The proof above makes it clear that the only children of a semigroup $\Lambda$ that may have infinitely many descendants are $\Lambda\setminus\{c(\Lambda)\}$ and $\Lambda\setminus\{c(\Lambda)+1\}$. 
\end{remark}

\begin{proposition}\label{proposition:infinite-child}
Except for $ \mathbb{N}_0$ and hyperelliptic semigroups, every numerical semigroup that is in an infinite chain has at least one child that is not. 
\end{proposition}

\begin{proof}
  Let $\Lambda$ be a numerical semigroup such that $\text{gcd}(  \mathcal{L}(\Lambda))\neq1$. Suppose that $\Lambda$ only has children that are in infinite chains. Let $c=c(\Lambda)$. By Corollary~\ref{corollary:infinite-chain}, the set of effective generators is either $\{c\}$, $\{c + 1\}$, or $\{c, c + 1\}$.

In the first case, the only child of $\Lambda$ is $\Lambda_1=\Lambda\setminus\{c\}$ and, since $\Lambda$ is in an infinite chain, $\Lambda_1$ cannot be a leaf. By Lemma~\ref{lemma:effective-generators}, $\Lambda_1$ has a (unique) effective generator which is $c+m$, where $m=m(\Lambda)$. So, the child of $\Lambda_1$ is $\Lambda_2=\Lambda_1\setminus\{c+m\}=\Lambda \setminus\{c,c+m\}$. If $m>2$ then, as $c$ is the conductor of $\Lambda$, we have $\{c+1,c+2\}\subset\Lambda\setminus\{c,c+m\}=\Lambda_2$. However, since $\Lambda_2$ is the only child of the only child of $\Lambda$, it must also be in an infinite chain. That is, gcd$(  \mathcal{L}(\Lambda_2))\neq 1$ but gcd$(c+1,c+2)=1$, a contradiction. 

Similarly, in the second case if the only effective generator of $\Lambda$ is $c+1$, then all its descendants are sticks, the next ones being $\Lambda_1=\Lambda\setminus\{c+1\}$ and $\Lambda_2=\Lambda_1\setminus\{c+1+m\}=\Lambda\setminus\{c+1,c+1+m\}$. And if $m>2$, $\{c+2,c+3\}\subset\Lambda_2$, contradicting gcd$(  \mathcal{L}(\Lambda_2))\neq1$.

Finally, for the third case we assume that $c$ and $c+1$ are the only effective generators of $\Lambda$. So, when we take out $c+1$ we are in the same situation as in the second case, where we need $m\leq2$. 

Therefore the only possibility for $\Lambda$ to have all children in infinite chains is if the multiplicity of $\Lambda$ is equal to $1$ or $2$, that is, if $\Lambda$ is $ \mathbb{N}_0$ or a hyperelliptic semigroup.
\end{proof}

A direct conclusion from Proposition~\ref{proposition:infinite-child} is that, with the exception of $ \mathbb{N}_0$ and hyperelliptic semigroups, a numerical semigroup that is in an infinite chain has at least two children.

The way the tree of numerical semigroups is constructed generates subtrees of numerical semigroups with the same multiplicity. We will comment more on this in the next section. For now, consider the following result.

\begin{lemma}\label{lemma:infinite-by-multiplicity}
For every pair $(g,m),g\geq1,2\leq m\leq g+1,$ there exists a numerical semigroup $\Lambda$ with $g(\Lambda)=g$ that lies in an infinite chain formed by numerical semigroups of multiplicity $m$. 
\end{lemma}
\begin{proof}
Consider the chain $\omega(m, {\mathbb N}_0)$ in the notation of Lemma~\ref{lem:semintegertoinfinitechain} and note that, starting from
$\{0, m, \rightarrow\}$, all semigroups in this chain have multiplicity $m$, and since they form an unbroken path in
the tree, they achieve all genera $g\geq  m - 1$.
\end{proof}

\begin{lemma}\label{lemma:quota-for-efficacy}
Let $\Lambda$ be a numerical semigroup with enumeration $\lambda$, genus $g$, conductor $c$, and gcd$(\mathcal{L}(\Lambda))=d>1$. Then the number of children of $\Lambda$ is at least $d-1$ and if $\lambda_1=d$, then the number of children of $\Lambda$ is exactly $d-1$.  

\end{lemma}

\begin{proof}
Since $d$ is a divisor of the left elements of $\Lambda$, we have $d\leq \lambda_1$, so $d-1<\lambda_1$.  Since $d\neq 1$, the set $\{c,c+1,\dots,c+d-1\}$ has more than one element and it has exactly one multiple of $d$. A non-multiple of $d$ of the form $c+i$ with $0\leq i\leq d-1$ cannot be generated by $\{\lambda_1,\dots,\lambda_{c-g-1}\}\cup\{c,c+1,\dots,c+i-1\}$. Indeed, let $  \mathcal{R}=\{c,c+1,\dots,c+i-1\}$. Suppose that $c+i=\lambda_{j_1}+\dots+\lambda_{j_k}$. But none of the $\lambda_{j_s}$ is in $  \mathcal{R}$ since, otherwise, if $\lambda_{j_s}\in \mathcal{R}$, then $\lambda_{j_s}\geq c$, and so $c+i-\lambda_{j_s}\leq i \leq d-1<\lambda_1$, a contradiction. Therefore, every numerical semigroup that is in an infinite chain has at least $d-1$ children. Moreover, if $\lambda_1=d$, then, by Proposition~\ref{proposition:most-effective-generator}, $\Lambda$ has exactly $d-1$ effective generators.
\end{proof}

\begin{definition}
A numerical semigroup is \textit{fertile} if most of its children are in infinite chains.
\end{definition}

\begin{proposition}\label{proposition:fertile-equivalence}
A numerical semigroup $\Lambda$ is fertile if and only if one of the options below holds:
\begin{enumerate} 
    \item $\Lambda= \mathbb{N}_0$;
    \item $\Lambda$ is a hyperelliptic semigroup;
    \item\label{problematic-case} $h(\Lambda)=3$ and two of its children are in infinite chains.
\end{enumerate}
\end{proposition}

\begin{proof}
Let $\Lambda$ be a fertile numerical semigroup. By Corollary~\ref{corollary:infinite-chain}, if $h(\Lambda)>3$ then it has at most $2$ children in infinite chains, so it cannot be fertile. If $h(\Lambda)=3$, then it is fertile if it has $2$ children in infinite chains, which corresponds to the case~\ref{problematic-case}. If $h(\Lambda)=2$, then $\Lambda$ is fertile if both its children are in infinite chains and, by Proposition~\ref{proposition:infinite-child}, that occurs when $\Lambda= \mathbb{N}_0\setminus\{1\}$, which is hyperelliptic as desired. Finally, if $h(\Lambda)=1$ and its only child is in an infinite chain, then $\Lambda$ is a hyperelliptic semigroup or $\mathbb{N}_0$. 
\end{proof}

There is one hyperelliptic semigroup at each level and we would like to explore what is the frequency of numerical semigroups of type~\ref{problematic-case} at each level of the semigroup tree. Consider the semigroup sequence defined by $M_n:=\{4k:k\geq0\}\cup [4n+2,\infty)$.  

Note that, if $\Lambda=M_n$ for some $n\geq 1$, then $c(\Lambda)=4n+2$ is a minimal generator of $\Lambda$, gcd$( \mathcal{L}(\Lambda))=$ gcd$(0,4,\dots,4n)=4$ and gcd$(\mathcal{L}(\Lambda \setminus \{c(\Lambda)+1\}))=$ gcd$(  \mathcal{L}(\Lambda),4n+2)=2$. So, by Theorem~\ref{theorem:infinitechains}, $\Lambda \setminus \{c\}$ and $\Lambda \setminus \{c+1\}$ are in infinite chains. In addition, $c(\Lambda)+3=4n+5$ is also an effective generator and, by Proposition~\ref{proposition:most-effective-generator}, it is the last of them. Therefore, $\Lambda=M_n$ is of type~\ref{problematic-case}.

\begin{proposition}\label{proposition:type-c-equivalence}
A numerical semigroup $\Lambda$ of genus $g>2$ is of type~\ref{problematic-case} if and only if $\Lambda=M_n$ for some $n\geq 1$, where $M_n=\{4k:k\geq0\}\cup [4n+2,\infty)$.
\end{proposition}

\begin{proof}
The only ordinary numerical semigroup which is of type~\ref{problematic-case} has genus $2$ since the number of effective generators of the ordinary semigroup of genus $g$ is $g+1$, while type~\ref{problematic-case} semigroups have efficacy equal to 3. Let $\Lambda$ be a numerical semigroup of type~\ref{problematic-case}, and let $g=g(\Lambda)>2,c=c(\Lambda)$ be its genus and conductor and let $\lambda$ be its enumeration. Then, gcd$(\lambda_0, \dots, \lambda_{c-g-1})=d>1$ and, since $h(\Lambda)=3$, by Lemma~\ref{lemma:quota-for-efficacy}, we have $d\leq 4$. Moreover, by Remark~\ref{remark:descendants-type-c}, $c$ and $c+1$ are generators of $\Lambda$, with gcd$(\lambda_0, \dots, \lambda_{c-g-1},\lambda_{c-g})=d'> 1$. 

If $d=2$, then $d'=2$, $c$ is even, and $\lambda_1\neq2$ by Lemma~\ref{lemma:quota-for-efficacy}. In this case, all nongaps before $c+1$ are even, so if $c+3$ is not a generator, $2\in \Lambda$, which is a contradiction. Thus $c+3$ is the largest effective generator and the only possibility to write $c+5$ as the sum of nongaps is $c+5=(c+1)+4$. Then, $4\in\Lambda$ and $c$ must be congruent to $2$ modulo $4$, in order to be a generator. Now, since $d\neq 4$, there must exist $\lambda_i\in \mathcal{L}(\Lambda)$ such that $4 \nmid \lambda_i$. That is, $\lambda_i$ is congruent to $2$ modulo $4$. But then $c-\lambda_i$ is a multiple of $4$, and so it belongs to $\Lambda$, contradicting the fact that $c$ is a generator.

If $d=3$, similarly we have $3\mid c$, and so $3\nmid c+2$. Thus, $c+2$ is not equal to the sum of any elements of $\Lambda$ smaller than $c$, so $c+2$ is the third effective generator. Consequently $c,c+1,c+2$ are the three effective generators of $\Lambda$ and, since $c+4$ is not multiple of $3$ and it is not a generator, writing $c+4$ as the sum of nongaps we get $c+4=(c+1)+3$ or $c+4=(c+2)+2$. So, either $2\in\Lambda$ or $3\in\Lambda$ but $2\in \Lambda$ contradicts $d=3$ and $3\in\Lambda$ contradicts $c$ being a generator. 

If $d=4$ and $m>4$, then $c$ is even because $d'\mid 4$ and $d'\neq 1$. Thus, $c+3$ and $c+5$ are odd, and so they are generators of $\Lambda$, contradicting $h(\Lambda)=3$. 

If $d=4$ and $m=4$, then all left elements are multiples of $4$, and $c$ needs to be even and not multiple of $4$ since we have $d'\neq 1$ and $c$ a generator. This description corresponds exactly to the semigroups of the sequence $M_n$. 
\end{proof}

It is easy to see that $g(M_n)=3n+1$. Thus we obtain the following result. 

\begin{theorem}\label{theorem:fertile}
The unique fertile semigroups of genus $g$ are:

\begin{enumerate}
\item ${  \mathbb N}_0$ if $g=0$.
\item ${  \mathbb N}_0\setminus\{1\}$ if $g=1$.
\item $\mathcal{H}_g$ if $g>1$ and $g\not\equiv 1 \mod 3$.
\item $\mathcal{H}_g$ and $M_{\frac{g-1}{3}}$ if $g>1$ and $g\equiv 1 \mod 3$.
\end{enumerate}

\end{theorem}

\begin{proof}
Direct from Propositions~\ref{proposition:fertile-equivalence} and~\ref{proposition:type-c-equivalence}.
\end{proof}

Let $i_g$ be the number of numerical semigroups of genus $g$ that are in an infinite chain. 

\begin{theorem}\label{theorem:majority}
For $g\geq 5$ we have $i_g\leq\dfrac{1}{2}n_g$.
\end{theorem}

\begin{proof}
Computations show that $i_5=6$ and $n_5=12$. For fixed $g\geq6$, let us prove that there are more numerical semigroups with genus $g$ that are not in infinite chains than the other way around. To do so, we partition the set of numerical semigroups of genus $g$ into three disjoint sets:
\begin{align*}
    F_g&:=\{\Lambda:g(\Lambda)=g \text{ and } \Lambda\cup\{F(\Lambda)\} \text{ is fertile}\}\\
    O_g&:=\{\Lambda:g(\Lambda)=g \text{ and } \Lambda\cup\{F(\Lambda)\} \text{ is ordinary}\}\\
    P_g&:=\{\Lambda:g(\Lambda)=g \text{ and } \Lambda\notin F_g\sqcup O_g\}
\end{align*}
By Lemma~\ref{lemma:infinite-by-multiplicity}, $P_g\neq \emptyset$. By Theorem~\ref{theorem:fertile} at most half the nodes in $P_g$ lie in an infinite chain.

If $g-1\equiv1\mod 3$ then, by Theorem~\ref{theorem:fertile}, there exist $3$ numerical semigroups in $F_g$ which are in an infinite chain and $1$ which is not. In $O_g$ there are $2$ elements in an infinite chain and $g-2\geq4$ that are not. In this way, the difference between the number of elements of $F_g\sqcup O_g$ which are not in an infinite chain and the number of those that are is non-negative as well. If $g-1\not\equiv1\mod 3$, then $F_g$ has only $1$ element and it belongs to an infinite chain and the result holds similarly. 
\end{proof}

\section{Fixing the multiplicity}\label{sec:multi}

In the tree that organizes all semigroups by genus, the set of descendants of $\mathbb{N}_0\setminus\{1,3\}$ is a single infinite chain of hyperelliptic semigroups.
Indeed, $\mathbb{N}_0\setminus\{1,3\}$ is the third semigroup in $I_{\mathcal H}$. On the other hand $\mathbb{N}_0\setminus\{1,2\}$ is in $I_{\mathcal O}$, the infinite chain formed by the ordinary semigroups. From each ordinary semigroup in $I_{\mathcal O}$ there emerges a subtree that contains all the semigroups of the same multiplicity (as one can see in \cite{m-towards}). From now on, we focus on these subtrees with fixed multiplicity. One way to transit in such a subtree is, from a semigroup in it, to move to another numerical semigroup of higher genus while maintaining the structure of the first semigroup, that is, maintaining the structure of its non-gaps. The construction explained in next definition does this: it pushes the non-gaps of a numerical semigroup while keeping its multiplicity.

\begin{definition}\label{def:push}
The {\it push} of a numerical semigroup $\Lambda$, with enumeration $\lambda$, {\it by its multiplicity} is $\lambda_1\oplus\Lambda:=\{0\} \cup \{\lambda_1+\lambda_j;\lambda_j\in \Lambda\}.$
\end{definition}

The set $\lambda_1\oplus\Lambda$ as above is a numerical semigroup. In fact, if $x,y\in\lambda_1\oplus\Lambda$ then, {either $x,y$ or both are zero, in which case the sum belongs to $\lambda_1\oplus\Lambda$, or} there exist $\lambda_a,\lambda_b\in\Lambda$ such that $x=\lambda_1+\lambda_a$ and $y=\lambda_1+\lambda_b$. {In this case,} $x+y=\lambda_1+(\lambda_a+\lambda_1+\lambda_b)\in \lambda_1\oplus\Lambda$. In addition, if $x\geq \lambda_1+c(\Lambda)$, then $x-\lambda_1\in\Lambda$, and so $x\in \lambda_1\oplus \Lambda$. That is, $\lambda_1+c(\Lambda)-1$ is the maximum of $(\lambda_1\oplus\Lambda)^C$ and, thus, $(\lambda_1\oplus\Lambda)^C$ is finite. Moreover, $m(\lambda_1\oplus\Lambda)=\lambda_1$, $g(\lambda_1\oplus\Lambda)=g(\Lambda)+\lambda_1-1$, $c(\lambda_1\oplus\Lambda)=\lambda_1+c(\Lambda)$, $\mathcal{L}(\lambda_1\oplus \Lambda)=\{0\}\cup \{\lambda_1+\lambda_j;\lambda_j\in  \mathcal{L}(\Lambda)\}$ and $\#  \mathcal{L}(\lambda_1\oplus \Lambda)=\#  \mathcal{L}(\Lambda)+1$. 

\begin{lemma}\label{lemma:push}
Let $\Lambda$ be a numerical semigroup with enumeration $\lambda$ and let $\Pi$ be a non-ordinary numerical semigroup. Then,
\begin{enumerate}
    \item\label{effective-generator-push} If $\lambda_k$ is an (effective) generator of $\Lambda$, then $\lambda_1+\lambda_k$ is an (effective) generator of $\lambda_1  \oplus \Lambda$.
    \item\label{child-preservation} $\Pi$ is a child of $\Lambda$ if and only if $\lambda_1\oplus \Pi$  is a child of $\lambda_1\oplus \Lambda$.
    \item\label{infinite-preservation} $\Lambda$ lies in an infinite chain if and only if so does $\lambda_1\oplus \Lambda$.
      \end{enumerate}
\end{lemma}
\begin{proof}
Direct by Definition~\ref{def:push}.
\end{proof}

The reciprocal of implication~\ref{effective-generator-push} is not true in general. For instance, $\langle4,5,6\rangle$ is a leaf, while $4\oplus \langle4,5,6\rangle$ is a stick. From now on, we use the notation $\lambda_1\oplus^n \Lambda$ to represent $\underbrace{\lambda_1\oplus (\cdots(\lambda_1\oplus(\lambda_1\oplus\Lambda)))}_{n \text{ times}}$.

Now we will investigate infinite chains in subtrees with some fixed multiplicity. We call {\it infinite chains of multiplicity $m$} those infinite chains whose first $m$ semigroups are the ordinary semigroups of multiplicities $1$ to $m$ and after that all semigroups in the chain have multiplicity $m$. Equivalently, an infinite chain is an infinite chain of multiplicity $m$ if and only if the maximum of the multiplicities of the semigroups in the chain is $m$.

\subsection{Prime multiplicity}

From Theorem~\ref{theorem:infinitechains} one can deduce that, for each prime multiplicity, there is only one infinite chain of that multiplicity. Indeed, this unique chain is $\omega(m,{\mathbb N}_0)$. Hence, we can state the following theorem.

\begin{theorem}\label{theorem:primem}
  If $m$ is a prime, then the number $i_g(m)$ of numerical semigroups of genus $g\geq m-1$ and multiplicity $m$ that are in an infinite chain is $1$.
\end{theorem}

\subsection{Multiplicity 4}\label{subsec-m4}
\begin{figure}[ht]
    \centering
    \caption{Tree structure of numerical semigroups of multiplicity $4$ that are in an infinite chain, from genus $3$ to $5$.}
    \label{fig:m4-subtree}

\begin{tikzpicture}[grow=right]
\node {$A$} [red] 
child{node[black] {$B$}
child {node[black]{$D$}}
child {node[black]{$C$}}}
;

\end{tikzpicture}
\end{figure}
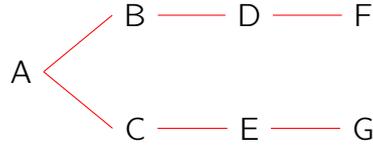

The tree in Figure~\ref{fig:m4-subtree} represents the numerical semigroups of multiplicity $4$, genus from $3$ to $5$ and which are in infinite chains. The edges connect semigroups as usual, being a child of a given semigroup the given semigroup minus one of its effective generators.

Specifically: $A=\{0,4,\rightarrow\}, B=\{0,4,6,\rightarrow\}, C=\{0,4,7,\rightarrow\}$, and $ D=\{0,4,6,8,\rightarrow\}$.

As we can observe, $c(C)$ is a generator while $c(C)+1$ is not. So by Remark \ref{remark:descendants-type-c} $C$ has only one child in an infinite chain and this child is exactly $4\oplus A$. In general, for $n\geq1$, $4\oplus^n C\setminus \{c(4\oplus^n C)\}=4\oplus^{n+1}A$ is the only child of $4\oplus^n C$ that is in an infinite chain. Moreover, $4\oplus^n A$ has one child in an infinite chain, and so by Lemma \ref{lemma:push} this child is $4\oplus^n B$.

On the other hand, the efficacy of $4\oplus^n D=\{0,4,\cdots,4n,4n+2,4n+4,\rightarrow\}$, with $n\geq1$, is equal to $2$. So the efficacy of any descendant of $4\oplus^n D$ is at most $2$. Then by Proposition~\ref{proposition:infinite-child} every descendant of $4\oplus^n D$ has exactly one child in an infinite chain. So, the infinite chain emerging from a node of the form $4\oplus^n D$ is made of sticks
in the tree of semigroups that are in infinite chains. In that way we have the following theorem.

\begin{theorem}\label{theorem:mfour}
  The number $i_g(4)$ of numerical semigroups of genus $g\geq 3$ and multiplicity $4$ that are in an infinite chain is
  $$\left\lfloor\dfrac{g+1}{3}\right\rfloor.$$
\end{theorem}

\begin{proof}
It is clear for $g=3,4$. Otherwise, the tree in Figure~\ref{fig:m4-subtree} is self-replicated in the tree of infinite chains of multiplicity $4$ starting at $\{0,4,\rightarrow\}$. Moreover, {as observed before,} the numerical semigroups of multiplicity $4$ that are in an infinite chain and do not participate in self-replication are sticks in the tree of semigroups that are in infinite chains. 
\end{proof}

\begin{example}
  In the tree in Figure~\ref{fig:example-m4-g40}
we can observe thirteen complete self-replications of the tree in Figure~\ref{fig:m4-subtree}.
The figure  was generated with the {\tt drawsgtree} tool \cite{drawsgtree}, by command \textsc{./drawsgtree -g41 -m4 -e infinitechains -etrim -d .2}. 
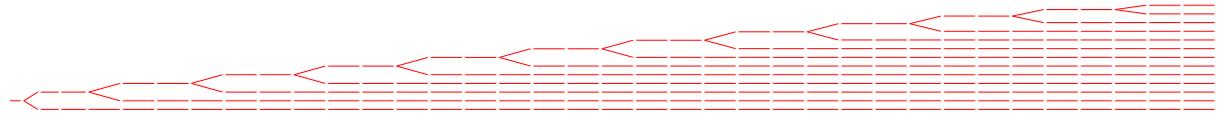
\begin{figure}[htbp]
    \centering
    \caption{Tree of numerical semigroups in infinite chains, with multiplicity $4$, from genus $3$ up to $41$.}
    \vspace{0.5cm}
    \label{fig:example-m4-g40}
\adjustbox{max width=\textwidth,max height=.9\textheight}{\begin{tikzpicture}[grow'=right, sibling distance=6.000000mm]\tikzset{level 1/.style={level distance=1.400000cm}}\tikzset{level 2/.style={level distance=1.750000cm}}\tikzset{level 3/.style={level distance=2.100000cm}}\tikzset{level 4/.style={level distance=2.800000cm}}\tikzset{level 5+/.style={level distance=3.500000cm}}\Tree[.{}  \edge [red]; [.{}  \edge [red]; [.{}  \edge [red]; [.{}  \edge [red]; [.{}  \edge [red]; [.{}  \edge [red]; [.{}  \edge [red]; [.{}  \edge [red]; [.{}  \edge [red]; [.{}  \edge [red]; [.{}  \edge [red]; [.{}  \edge [red]; [.{}  \edge [red]; [.{}  \edge [red]; [.{}  \edge [red]; [.{}  \edge [red]; [.{}  \edge [red]; [.{}  \edge [red]; [.{}  \edge [red]; [.{}  \edge [red]; [.{}  \edge [red]; [.{}  \edge [red]; [.{}  \edge [red]; [.{}  \edge [red]; [.{}  \edge [red]; [.{}  \edge [red]; [.{}  \edge [red]; [.{}  \edge [red]; [.{}  \edge [red]; [.{}  \edge [red]; [.{}  \edge [red]; [.{}  \edge [red]; [.{}  \edge [red]; [.{}  \edge [red]; [.{}  \edge [red]; [.{}  \edge [red]; [.{}  \edge [red]; [.{}  \edge [red]; [.{} ] \edge [red]; [.{} ]]]] \edge [red]; [.{}  \edge [red]; [.{}  \edge [red]; [.{}  \edge [red]; [.{} ]]]]]]] \edge [red]; [.{}  \edge [red]; [.{}  \edge [red]; [.{}  \edge [red]; [.{}  \edge [red]; [.{}  \edge [red]; [.{}  \edge [red]; [.{} ]]]]]]]]]] \edge [red]; [.{}  \edge [red]; [.{}  \edge [red]; [.{}  \edge [red]; [.{}  \edge [red]; [.{}  \edge [red]; [.{}  \edge [red]; [.{}  \edge [red]; [.{}  \edge [red]; [.{}  \edge [red]; [.{} ]]]]]]]]]]]]] \edge [red]; [.{}  \edge [red]; [.{}  \edge [red]; [.{}  \edge [red]; [.{}  \edge [red]; [.{}  \edge [red]; [.{}  \edge [red]; [.{}  \edge [red]; [.{}  \edge [red]; [.{}  \edge [red]; [.{}  \edge [red]; [.{}  \edge [red]; [.{}  \edge [red]; [.{} ]]]]]]]]]]]]]]]] \edge [red]; [.{}  \edge [red]; [.{}  \edge [red]; [.{}  \edge [red]; [.{}  \edge [red]; [.{}  \edge [red]; [.{}  \edge [red]; [.{}  \edge [red]; [.{}  \edge [red]; [.{}  \edge [red]; [.{}  \edge [red]; [.{}  \edge [red]; [.{}  \edge [red]; [.{}  \edge [red]; [.{}  \edge [red]; [.{}  \edge [red]; [.{} ]]]]]]]]]]]]]]]]]]] \edge [red]; [.{}  \edge [red]; [.{}  \edge [red]; [.{}  \edge [red]; [.{}  \edge [red]; [.{}  \edge [red]; [.{}  \edge [red]; [.{}  \edge [red]; [.{}  \edge [red]; [.{}  \edge [red]; [.{}  \edge [red]; [.{}  \edge [red]; [.{}  \edge [red]; [.{}  \edge [red]; [.{}  \edge [red]; [.{}  \edge [red]; [.{}  \edge [red]; [.{}  \edge [red]; [.{}  \edge [red]; [.{} ]]]]]]]]]]]]]]]]]]]]]] \edge [red]; [.{}  \edge [red]; [.{}  \edge [red]; [.{}  \edge [red]; [.{}  \edge [red]; [.{}  \edge [red]; [.{}  \edge [red]; [.{}  \edge [red]; [.{}  \edge [red]; [.{}  \edge [red]; [.{}  \edge [red]; [.{}  \edge [red]; [.{}  \edge [red]; [.{}  \edge [red]; [.{}  \edge [red]; [.{}  \edge [red]; [.{}  \edge [red]; [.{}  \edge [red]; [.{}  \edge [red]; [.{}  \edge [red]; [.{}  \edge [red]; [.{}  \edge [red]; [.{} ]]]]]]]]]]]]]]]]]]]]]]]]] \edge [red]; [.{}  \edge [red]; [.{}  \edge [red]; [.{}  \edge [red]; [.{}  \edge [red]; [.{}  \edge [red]; [.{}  \edge [red]; [.{}  \edge [red]; [.{}  \edge [red]; [.{}  \edge [red]; [.{}  \edge [red]; [.{}  \edge [red]; [.{}  \edge [red]; [.{}  \edge [red]; [.{}  \edge [red]; [.{}  \edge [red]; [.{}  \edge [red]; [.{}  \edge [red]; [.{}  \edge [red]; [.{}  \edge [red]; [.{}  \edge [red]; [.{}  \edge [red]; [.{}  \edge [red]; [.{}  \edge [red]; [.{}  \edge [red]; [.{} ]]]]]]]]]]]]]]]]]]]]]]]]]]]] \edge [red]; [.{}  \edge [red]; [.{}  \edge [red]; [.{}  \edge [red]; [.{}  \edge [red]; [.{}  \edge [red]; [.{}  \edge [red]; [.{}  \edge [red]; [.{}  \edge [red]; [.{}  \edge [red]; [.{}  \edge [red]; [.{}  \edge [red]; [.{}  \edge [red]; [.{}  \edge [red]; [.{}  \edge [red]; [.{}  \edge [red]; [.{}  \edge [red]; [.{}  \edge [red]; [.{}  \edge [red]; [.{}  \edge [red]; [.{}  \edge [red]; [.{}  \edge [red]; [.{}  \edge [red]; [.{}  \edge [red]; [.{}  \edge [red]; [.{}  \edge [red]; [.{}  \edge [red]; [.{}  \edge [red]; [.{} ]]]]]]]]]]]]]]]]]]]]]]]]]]]]]]] \edge [red]; [.{}  \edge [red]; [.{}  \edge [red]; [.{}  \edge [red]; [.{}  \edge [red]; [.{}  \edge [red]; [.{}  \edge [red]; [.{}  \edge [red]; [.{}  \edge [red]; [.{}  \edge [red]; [.{}  \edge [red]; [.{}  \edge [red]; [.{}  \edge [red]; [.{}  \edge [red]; [.{}  \edge [red]; [.{}  \edge [red]; [.{}  \edge [red]; [.{}  \edge [red]; [.{}  \edge [red]; [.{}  \edge [red]; [.{}  \edge [red]; [.{}  \edge [red]; [.{}  \edge [red]; [.{}  \edge [red]; [.{}  \edge [red]; [.{}  \edge [red]; [.{}  \edge [red]; [.{}  \edge [red]; [.{}  \edge [red]; [.{}  \edge [red]; [.{}  \edge [red]; [.{} ]]]]]]]]]]]]]]]]]]]]]]]]]]]]]]]]]] \edge [red]; [.{}  \edge [red]; [.{}  \edge [red]; [.{}  \edge [red]; [.{}  \edge [red]; [.{}  \edge [red]; [.{}  \edge [red]; [.{}  \edge [red]; [.{}  \edge [red]; [.{}  \edge [red]; [.{}  \edge [red]; [.{}  \edge [red]; [.{}  \edge [red]; [.{}  \edge [red]; [.{}  \edge [red]; [.{}  \edge [red]; [.{}  \edge [red]; [.{}  \edge [red]; [.{}  \edge [red]; [.{}  \edge [red]; [.{}  \edge [red]; [.{}  \edge [red]; [.{}  \edge [red]; [.{}  \edge [red]; [.{}  \edge [red]; [.{}  \edge [red]; [.{}  \edge [red]; [.{}  \edge [red]; [.{}  \edge [red]; [.{}  \edge [red]; [.{}  \edge [red]; [.{}  \edge [red]; [.{}  \edge [red]; [.{}  \edge [red]; [.{} ]]]]]]]]]]]]]]]]]]]]]]]]]]]]]]]]]]]]] \edge [red]; [.{}  \edge [red]; [.{}  \edge [red]; [.{}  \edge [red]; [.{}  \edge [red]; [.{}  \edge [red]; [.{}  \edge [red]; [.{}  \edge [red]; [.{}  \edge [red]; [.{}  \edge [red]; [.{}  \edge [red]; [.{}  \edge [red]; [.{}  \edge [red]; [.{}  \edge [red]; [.{}  \edge [red]; [.{}  \edge [red]; [.{}  \edge [red]; [.{}  \edge [red]; [.{}  \edge [red]; [.{}  \edge [red]; [.{}  \edge [red]; [.{}  \edge [red]; [.{}  \edge [red]; [.{}  \edge [red]; [.{}  \edge [red]; [.{}  \edge [red]; [.{}  \edge [red]; [.{}  \edge [red]; [.{}  \edge [red]; [.{}  \edge [red]; [.{}  \edge [red]; [.{}  \edge [red]; [.{}  \edge [red]; [.{}  \edge [red]; [.{}  \edge [red]; [.{}  \edge [red]; [.{}  \edge [red]; [.{} ]]]]]]]]]]]]]]]]]]]]]]]]]]]]]]]]]]]]]]]\end{tikzpicture}}

\end{figure}
\end{example}

\subsection{Multiplicity 6}\label{subsec-m6}
The numerical semigroup of multiplicity $6$ and with the smallest genus is the ordinary semigroup $\{0,6,\rightarrow\}$. It has two children in infinite chains: the ordinary semigroup $\{0,7,\rightarrow\}$ and the semigroup $\{0,6,8,\rightarrow\}$. To study the numerical semigroups that are in infinite chains and have multiplicity $6$, we study the tree $\upsilon$ formed by $\{0,6,\rightarrow\}$ in its root and connected to 
$\{0,6,8,\rightarrow\}$ and all semigroups descending from it 
that are in infinite chains. In other words, $\upsilon$ is formed by the subtree rooted at $\{0,6,\rightarrow\}$ that contains all numerical semigroups in infinite chains, where the node $\{0,7,\rightarrow\}$ and all its descendants have been trimmed.

The semigroups of $\upsilon$ are of type $\Lambda_n=\{0,6,\dots,6n,6n+\beta_1,\dots,6n+\beta_j,\rightarrow\}$. In this case we can use the $j$-tuple $(\beta_1\dots,\beta_j)_n$ to represent $\Lambda_n$. For example, we represent $\{0,6,8,\rightarrow\}$ by $(2)_1$ and $\{0,6,12,14,16,18,20,22,24,\rightarrow\}$ by $(2,4,6,8,10,12)_2$. We use $(\cdot)_n$ to represent the ordinary semigroup of genus $6n-1$.

\subsubsection[The tree tau and its replications]{The tree $\tau_n$ and its replications}
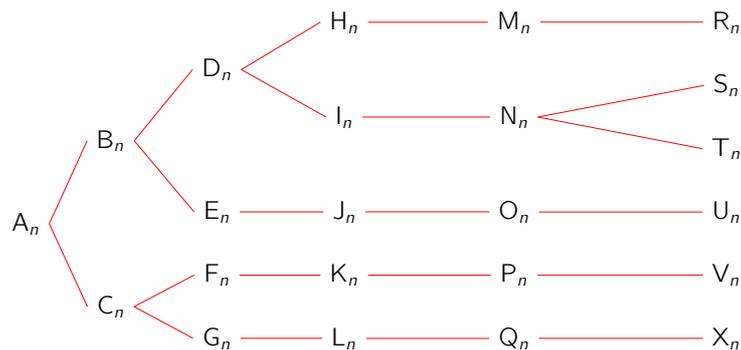
\begin{figure}[!ht]
    \caption{Finite tree $\tau_n$}
    \vspace{0.5cm}
    \label{fig:tree-m6}
    \centering
\adjustbox{max width=\textwidth,max height=.9\textheight}{\begin{tikzpicture}[grow'=right, sibling distance=6.000000mm]\tikzset{level 1/.style={level distance=1.400000cm}}\tikzset{level 2/.style={level distance=1.750000cm}}\tikzset{level 3/.style={level distance=2.100000cm}}\tikzset{level 4/.style={level distance=2.800000cm}}\tikzset{level 5+/.style={level distance=3.500000cm}}\Tree[.{$A_n$}  \edge [red]; [.{$B_n$}  \edge [red]; [.{$C_n$}  \edge [red]; [.{$E_n$}  \edge [red]; [.{$I_n$} ] \edge [red]; [.{$J_n$} ]] \edge [red]; [.{$F_n$}  \edge [red]; [.{$K_n$} ]]] \edge [red]; [.{$D_n$}  \edge [red]; [.{$G_n$}  \edge [red]; [.{$L_n$} ]] \edge [red]; [.{$H_n$}  \edge [red]; [.{$M_n$} ]]]]]\end{tikzpicture}}
\end{figure}

In Table~\ref{table-m6} we have thirteen numerical semigroups that have multiplicity $6$, are in infinite chains and have some parenting relationships that can be observed in the tree $\tau_n$ in Figure~\ref{fig:tree-m6}. By Remark~\ref{remark:descendants-type-c} and the last column in Table~\ref{table-m6}, we verify that each semigroup $\Lambda_n$ in $\tau_n$ has children in infinite chains as drawn in Figure~\ref{fig:m4-subtree}.

\begin{table}[ht]
\centering
\caption{Some numerical semigroups that have multiplicity $6$ and are in infinite chains.}
\label{table-m6}\vspace{0.5cm}
\begin{tabular}{|l|l|l|}
\hline
Label & Numerical Semigroup & Only children \\ 
\hline                               
$A_n$ & $(\cdot)_n$ &  $c(A_n)=6n$ is not a minimal generator\\
$B_n$ & $(2)_n$ &  \\
$C_n$ & $(3)_n$ &  \\
$D_n$ & $(2,4)_n$ &  \\
$E_n$ & $(4)_n$ &   \\
$F_n$ & $(3,5)_n$ & $c(F_n)+1=6n+6$ is not a minimal generator  \\
$G_n$ & $(2,5)_n$ & $c(G_n)+1=6n+6$ is not a minimal generator\\
$H_n$ & $(2,4,6)_n$ & $c(H_n)=6n+6$ is not a minimal generator\\
$I_n$ & $(5)_n$ & $c(I_n)+1=6n+6$ is not a minimal generator\\
$J_n$ & $(4,6)_n$ & \\
$K_n$ & $(3,6)_n$ & \\
$L_n$ & $(2,6)_n$ & \\
$M_n$ & $(2,4,6,8)_n$ & \\
\hline
\end{tabular}
\end{table}

%
%
%
%

Note that $I_n$ has only one child in an infinite chain which is $I_n\setminus\{c(I_n)\}=6\oplus A_n$. Moreover, $6\oplus \Lambda_n=\Lambda_{n+1}$ for all $n\geq 1$ and $\Lambda_n$ a numerical semigroup from Table~\ref{table-m6}.  Therefore, the finite tree in Figure~\ref{fig:tree-m6} is self-replicated in $\upsilon$ through push by multiplicity.

\subsubsection[Outside the replications of tau]{Outside the replications of $\tau_n$}
Now we focus on the numerical semigroups of multiplicity $6$ that are in an infinite chain and do not participate in that self-replication, i.e., the descendants of $J_n, K_n, L_n$, and $M_n$.

\begin{proposition}\label{prop:descandants-Tn-Xn}
  The numerical semigroups $K_n=\{0,6,\dots,6n,6n+3,6n+6,\rightarrow\}$ and
  $M_n=\{0,6,\dots,6n,6n+2,6n+4,6n+6,6n+8,\rightarrow\}$ each belong to one infinite chain. 
\end{proposition}

\begin{proof}
By Theorem~\ref{theorem:infinitechains} the number of infinite chains to which $K_n$ belongs is one plus the number of descendants of $\tilde K_n=\{0,2,\dots,2n,2n+1,2n+2,\rightarrow\}$ beyond $2n+1$. But $\tilde K_n$ has no generators larger than $2n+1$, and so it has no descendants beyond $2n+1$. Thus, $K_n$ only belongs to one infinite chain.

Similarly, the number of infinite chains to which $M_n$ belongs is one plus the number of descendants of $\tilde M_n=\{0,3,\dots,3n,3n+1,3n+2,3n+3,3n+4,\rightarrow\}$ beyond $3n+3$. But $\tilde M_n$ has no generators larger than $3n+3$, and so it has no descendants beyond $3n+3$. Thus, $M_n$ only belongs to one infinite chain.
\end{proof}

\begin{proposition}
The numerical semigroups $J_n=\{0,6,\dots,6n,6n+4,6n+6,\rightarrow\}$ belong to $n+1$ infinite chains. 
\end{proposition}

\begin{proof}
By Theorem~\ref{theorem:infinitechains}, the number of infinite chains in which $J_n$ lies is one plus the number of descendants of $\tilde J_n=\{0,3,\dots,3n,3n+2,3n+3,\rightarrow\}$ beyond $3n+2$.
  
If $n=1$, $\tilde J_n$ has one descendant beyond $5$. Hence, $J_1$ lies in two infinite chains.

Suppose $n>1$. Observe that $\tilde J_n$ has one unique child that contains $\{0,3,\dots,3n,3n+2\}$, which is $\tilde J_n\setminus\{3n+4\}$. Observe also that any semigroup containing $\{0,3,\dots,3n,3n+2\}$ must contain the semigroup $$\{0,3,\dots,3n,3n+1,3(n+1),3(n+1)+2\dots,3(2n),3(2n)+2,\rightarrow\}.$$ One can deduce that all descendants of $\tilde J_n$ beyond $3n+2$ are exactly the semigroups $\tilde J_{n,t}=\{0,3,\dots,3n,3n+2,3(n+1),3(n+1)+2,\dots,3(n+t),3(n+t)+2,\rightarrow\}$ for $2\leq t\leq n+1$. Hence, $J_n$ lies in exactly $n+1$ infinite chains.
\end{proof}

\begin{proposition}\label{proposition:n-descendants}
The numerical semigroups $L_n=\{0,6,\dots,6n,6n+2,6n+6,\rightarrow\}$ belong to $n$ infinite chains. 
\end{proposition}

\begin{proof}

By Theorem~\ref{theorem:infinitechains}, the number of infinite chains in which $L_n$ lies is one plus the number of descendants of $\tilde L_n=\{0,3,\dots,3n,3n+1,3n+3,\rightarrow\}$ beyond $3n+1$.
  
If $n=1$, $\tilde L_n$ has no descendants beyond $4$. Hence, $L_1$ lies in exactly one infinite chain.

Suppose $n>1$. Observe that $\tilde L_n$ has one unique child, which is $\tilde L_n\setminus\{3n+4\}$. Observe also that any semigroup containing $\{0,3,\dots,3n,3n+1\}$ must contain the semigroup $$\{0,3,\dots,3n,3n+1,3(n+1),3(n+1)+1\dots,3(2n),3(2n)+1,3(2n)+2,\rightarrow\}.$$ One can deduce that all descendants of $\tilde L_n$ beyond $3n+1$ are exactly the semigroups $\tilde L_{n,t}=\{0,3,\dots,3n,3n+1,3(n+1),3(n+1)+1,\dots,3(n+t),3(n+t)+1,\rightarrow\}$ for $2\leq t\leq n$. Hence, $L_n$ lies in exactly $n$ infinite chains.
\end{proof}

In order to know at which levels of the tree of numerical semigroups the descendants of the semigroups $J_n$ and $L_n$ has two children in infinite chains, let us define in the next two lemmas some numerical semigroups.

\begin{lemma}\label{lemma:gamma}
For every integer $t$ with $1\leq t \leq n$, the set \begin{eqnarray*}\gamma_{n,t}&=&\{0,6,\dots,6n,6n+4,7n,7n+4,\dots,6(n+t-1),6(n+t-1)+4,\\&&\phantom{mmmmmmmmmmmmmmmmm}6(n+t),6(n+t)+2,\rightarrow\}\\
      &=&\{0,6,\dots,6n,6n+a_1,6n+a_2,\dots,6n+a_{2t+1},\rightarrow\},
    \end{eqnarray*}
    with
$$a_i=\begin{cases}
3i+1 &\text{if } i \text{ is odd and } i\neq2t+1 \\
3i, &\text{if } i \text{ is even} \\
3i-1, &\text{if } i=2t+1. 
\end{cases}$$
    satisfies:
    \begin{enumerate}
    \item $\gamma_{n,t}$ is a numerical semigroup.
    \item $\gamma_{n,1}$ is a child of $J_n$ and, for $t\geq 2$, $\gamma_{n,t}$ is a descendant of $J_n$.
    \item If $t\leq n$, then $\gamma_{n,t}$ has two children in infinite chains.
    \item The semigroups $J_n$, $\gamma_{n,1},\gamma_{n,2}$, \dots,$\gamma_{n,n}$ are in the same infinite chain.
    \item The genus of $\gamma_{n,t}$ is $5n+4t+1$.
    \end{enumerate}
  \end{lemma}

The proof is straightforward and is left to the reader.

Let $\zeta_n$ be the infinite chain where all the semigroups $\{J_n,\gamma_{n,t}\}_{1\leq t\leq n}$ of Lemma~\ref{lemma:gamma} belong. Since each $\gamma_{n,t}$ has two children in infinite chains, one child belongs to $\zeta_n$ and the other one belongs to a different infinite chain. Thus, the $n+1$ infinite chains from Proposition~\ref{proposition:n-descendants} to which $S_n$ belongs are $\zeta_n$ and $n$ other chains, each of them containing $\gamma_{t,n}\setminus \{c(\gamma_{t,n})+1\}$, for $t$ between $1$ and $n$.

\begin{lemma}\label{lemma:nu}
For every integer $t$ with $1\leq t \leq n-1$, the set 
    \begin{eqnarray*}\nu_{n,t}&=&\{0,6,\dots,6n,6n+2,7n,7n+2,\dots,6(n+t-1),6(n+t-1)+2,\\&&\phantom{mmmmmmmmmmmmmmmmm}
      6(n+t),6(n+t)+2,6(n+t)+4\rightarrow\}\\
      &=&\{0,6,\dots,6n,6n+b_1,6n+b_2,\dots,6n+b_{2t+2},\rightarrow\},
    \end{eqnarray*}
    with
$$b_i=\begin{cases}
3i-1 &\text{if } i \text{ is odd}\\
3i, &\text{if } i \text{ is even and} i\neq 2t+2\\
3i-2, &\text{if } i=2t+2. 
\end{cases}$$
    satisfies:
    \begin{enumerate}
    \item $\nu_{n,t}$ is a numerical semigroup.
    \item $\nu_{n,1}$ is a grandchild of $L_n$ and, for $t\geq 2$, $\nu_{n,t}$ is a descendant of $L_n$.
    \item If $t\leq n-1$, then $\nu_{n,t}$ has two children in infinite chains.
    \item The semigroups $L_n,\nu_{n,1},\nu_{n,2}$,\dots,$\nu_{n,n-1}$ are in the same infinite chain.
      \item The genus of $\nu_{n,t}$ is $5n+4t+2$.
      \end{enumerate}
  \end{lemma}

Again, the proof is straightforward and is left to the reader.
  
Let $\eta_n$ be the infinite chain where all the semigroups $\{L_n,\nu_{n,t}\}_{1\leq t\leq n-1}$ of Lemma~\ref{lemma:nu} belong. For $n$ fixed, since each $\nu_{n,t}$ have two children in infinite chains, one child belongs to $\eta_n$ and the other one belongs to a different infinite chain. Thus, the $n$ infinite chains from Proposition~\ref{proposition:n-descendants} to which $L_n$ belongs are $\eta_n$ and $n-1$ other chains, each one containing $\nu_{t,n}\setminus \{c(\nu_{t,n})+1\}$, for $t$ between $1$ and $n-1$.

\subsubsection[Growth of the tree upsilon]{Growth of the tree $\upsilon$}
\begin{figure}[!htbp]
    \centering
    \caption{Numerical semigroups of multiplicity $6$ in infinite chains, from genus $5$ up to genus $29$. The subtree $\tau_3$ is highlighted in blue. The parts of the chains $\eta_3$ and $\zeta_3$ of genus larger than or equal to $19$ are highlighted in green.}
    \label{fig:example-m6-tau3-g31}
\adjustbox{max width=\textwidth,max height=.9\textheight}{\begin{tikzpicture}[grow'=right, sibling distance=6.000000mm]\tikzset{level 1/.style={level distance=1.700000cm}}\tikzset{level 2/.style={level distance=1.750000cm}}\tikzset{level 3/.style={level distance=1.050000cm}}\tikzset{level 4/.style={level distance=1.400000cm}}\tikzset{level 5+/.style={level distance=1.750000cm}}\Tree[.{}  \edge [red]; [.{}  \edge [red]; [.{}  \edge [red]; [.{}  \edge [red]; [.{}  \edge [red]; [.{}  \edge [red]; [.{}  \edge [red]; [.{}  \edge [red]; [.{}  \edge [red]; [.{}  \edge [red]; [.{$A_3$}  \edge [blue]; [.{$B_3$}  \edge [blue]; [.{$C_3$}  \edge [blue]; [.{$E_3$}  \edge [blue]; [.{$I_3$}  \edge [red]; [.{}  \edge [red]; [.{}  \edge [red]; [.{}  \edge [red]; [.{}  \edge [red]; [.{}  \edge [red]; [.{}  \edge [red]; [.{}  \edge [red]; [.{}  \edge [red]; [.{}  \edge [red]; [.{} ] \edge [red]; [.{} ]] \edge [red]; [.{}  \edge [red]; [.{} ]]] \edge [red]; [.{}  \edge [red]; [.{}  \edge [red]; [.{} ]] \edge [red]; [.{}  \edge [red]; [.{} ]]]]]] \edge [red]; [.{}  \edge [red]; [.{}  \edge [red]; [.{}  \edge [red]; [.{}  \edge [red]; [.{}  \edge [red]; [.{} ]]]] \edge [red]; [.{}  \edge [red]; [.{}  \edge [red]; [.{}  \edge [red]; [.{} ]]]]]]] \edge [red]; [.{}  \edge [red]; [.{}  \edge [red]; [.{}  \edge [red]; [.{}  \edge [red]; [.{}  \edge [red]; [.{}  \edge [red]; [.{} ]]]]]]]] \edge [red]; [.{}  \edge [red]; [.{}  \edge [red]; [.{}  \edge [red]; [.{}  \edge [red]; [.{}  \edge [red]; [.{}  \edge [red]; [.{}  \edge [red]; [.{} ]]] \edge [red]; [.{}  \edge [red]; [.{}  \edge [red]; [.{} ]]]]]]] \edge [red]; [.{}  \edge [red]; [.{}  \edge [red]; [.{}  \edge [red]; [.{}  \edge [red]; [.{}  \edge [red]; [.{}  \edge [red]; [.{} ]]]]]]]]]]] \edge [blue]; [.{$J_3$}  \edge [green]; [.{$\gamma_{3,1}$}  \edge [green]; [.{}  \edge [green]; [.{}  \edge [green]; [.{}  \edge [green]; [.{$\gamma_{3,2}$}  \edge [green]; [.{}  \edge [green]; [.{}  \edge [green]; [.{}  \edge [green]; [.{$\gamma_{3,3}$}  \edge [green]; [.{} ] \edge [green]; [.{} ]]]]] \edge [red]; [.{}  \edge [red]; [.{}  \edge [red]; [.{}  \edge [red]; [.{}  \edge [red]; [.{} ]]]]]]]]] \edge [red]; [.{}  \edge [red]; [.{}  \edge [red]; [.{}  \edge [red]; [.{}  \edge [red]; [.{}  \edge [red]; [.{}  \edge [red]; [.{}  \edge [red]; [.{}  \edge [red]; [.{} ]]]]]]]]]]]] \edge [blue]; [.{$F_3$}  \edge [blue]; [.{$K_3$}  \edge [red]; [.{}  \edge [red]; [.{}  \edge [red]; [.{}  \edge [red]; [.{}  \edge [red]; [.{}  \edge [red]; [.{}  \edge [red]; [.{}  \edge [red]; [.{}  \edge [red]; [.{}  \edge [red]; [.{} ]]]]]]]]]]]]] \edge [blue]; [.{$D_3$}  \edge [blue]; [.{$G_3$}  \edge [blue]; [.{$L_3$}  \edge [green]; [.{}  \edge [green]; [.{$\nu_{3,1}$}  \edge [green]; [.{}  \edge [green]; [.{}  \edge [green]; [.{}  \edge [green]; [.{$\nu_{3,2}$}  \edge [green]; [.{}  \edge [green]; [.{}  \edge [green]; [.{}  \edge [green]; [.{} ]]]] \edge [green]; [.{}  \edge [green]; [.{}  \edge [green]; [.{}  \edge [green]; [.{} ]]]]]]]] \edge [red]; [.{}  \edge [red]; [.{}  \edge [red]; [.{}  \edge [red]; [.{}  \edge [red]; [.{}  \edge [red]; [.{}  \edge [red]; [.{}  \edge [red]; [.{} ]]]]]]]]]]]] \edge [blue]; [.{$H_3$}  \edge [blue]; [.{$M_3$}  \edge [red]; [.{}  \edge [red]; [.{}  \edge [red]; [.{}  \edge [red]; [.{}  \edge [red]; [.{}  \edge [red]; [.{}  \edge [red]; [.{}  \edge [red]; [.{}  \edge [red]; [.{}  \edge [red]; [.{} ]]]]]]]]]]]]]]]] \edge [red]; [.{}  \edge [red]; [.{}  \edge [red]; [.{}  \edge [red]; [.{}  \edge [red]; [.{}  \edge [red]; [.{}  \edge [red]; [.{}  \edge [red]; [.{}  \edge [red]; [.{}  \edge [red]; [.{}  \edge [red]; [.{}  \edge [red]; [.{}  \edge [red]; [.{}  \edge [red]; [.{}  \edge [red]; [.{}  \edge [red]; [.{} ]]]]]]]]]] \edge [red]; [.{}  \edge [red]; [.{}  \edge [red]; [.{}  \edge [red]; [.{}  \edge [red]; [.{}  \edge [red]; [.{}  \edge [red]; [.{}  \edge [red]; [.{}  \edge [red]; [.{}  \edge [red]; [.{} ]]]]]]]]]]]]]] \edge [red]; [.{}  \edge [red]; [.{}  \edge [red]; [.{}  \edge [red]; [.{}  \edge [red]; [.{}  \edge [red]; [.{}  \edge [red]; [.{}  \edge [red]; [.{}  \edge [red]; [.{}  \edge [red]; [.{}  \edge [red]; [.{}  \edge [red]; [.{}  \edge [red]; [.{}  \edge [red]; [.{} ]]]]]]]]]]]]]]]]] \edge [red]; [.{}  \edge [red]; [.{}  \edge [red]; [.{}  \edge [red]; [.{}  \edge [red]; [.{}  \edge [red]; [.{}  \edge [red]; [.{}  \edge [red]; [.{}  \edge [red]; [.{}  \edge [red]; [.{}  \edge [red]; [.{}  \edge [red]; [.{}  \edge [red]; [.{}  \edge [red]; [.{}  \edge [red]; [.{}  \edge [red]; [.{}  \edge [red]; [.{} ]]]]]]]]]]]]]]]]]] \edge [red]; [.{}  \edge [red]; [.{}  \edge [red]; [.{}  \edge [red]; [.{}  \edge [red]; [.{}  \edge [red]; [.{}  \edge [red]; [.{}  \edge [red]; [.{}  \edge [red]; [.{}  \edge [red]; [.{}  \edge [red]; [.{}  \edge [red]; [.{}  \edge [red]; [.{}  \edge [red]; [.{}  \edge [red]; [.{}  \edge [red]; [.{}  \edge [red]; [.{}  \edge [red]; [.{} ]]]]]]]]]]]]] \edge [red]; [.{}  \edge [red]; [.{}  \edge [red]; [.{}  \edge [red]; [.{}  \edge [red]; [.{}  \edge [red]; [.{}  \edge [red]; [.{}  \edge [red]; [.{}  \edge [red]; [.{}  \edge [red]; [.{}  \edge [red]; [.{}  \edge [red]; [.{}  \edge [red]; [.{} ]]]]]]]]]]]]]]]]] \edge [red]; [.{}  \edge [red]; [.{}  \edge [red]; [.{}  \edge [red]; [.{}  \edge [red]; [.{}  \edge [red]; [.{}  \edge [red]; [.{}  \edge [red]; [.{}  \edge [red]; [.{}  \edge [red]; [.{}  \edge [red]; [.{}  \edge [red]; [.{}  \edge [red]; [.{}  \edge [red]; [.{}  \edge [red]; [.{}  \edge [red]; [.{}  \edge [red]; [.{} ]]]]]]]]]]]]]]]]]]]]] \edge [red]; [.{}  \edge [red]; [.{}  \edge [red]; [.{}  \edge [red]; [.{}  \edge [red]; [.{}  \edge [red]; [.{}  \edge [red]; [.{}  \edge [red]; [.{}  \edge [red]; [.{}  \edge [red]; [.{}  \edge [red]; [.{}  \edge [red]; [.{}  \edge [red]; [.{}  \edge [red]; [.{}  \edge [red]; [.{}  \edge [red]; [.{}  \edge [red]; [.{}  \edge [red]; [.{}  \edge [red]; [.{}  \edge [red]; [.{}  \edge [red]; [.{} ]]]]]]]]]]]]]]]]]]] \edge [red]; [.{}  \edge [red]; [.{}  \edge [red]; [.{}  \edge [red]; [.{}  \edge [red]; [.{}  \edge [red]; [.{}  \edge [red]; [.{}  \edge [red]; [.{}  \edge [red]; [.{}  \edge [red]; [.{}  \edge [red]; [.{}  \edge [red]; [.{}  \edge [red]; [.{}  \edge [red]; [.{}  \edge [red]; [.{}  \edge [red]; [.{}  \edge [red]; [.{}  \edge [red]; [.{}  \edge [red]; [.{} ]]]]]]]]]]]]]]]]]]]]]] \edge [red]; [.{}  \edge [red]; [.{}  \edge [red]; [.{}  \edge [red]; [.{}  \edge [red]; [.{}  \edge [red]; [.{}  \edge [red]; [.{}  \edge [red]; [.{}  \edge [red]; [.{}  \edge [red]; [.{}  \edge [red]; [.{}  \edge [red]; [.{}  \edge [red]; [.{}  \edge [red]; [.{}  \edge [red]; [.{}  \edge [red]; [.{}  \edge [red]; [.{}  \edge [red]; [.{}  \edge [red]; [.{}  \edge [red]; [.{}  \edge [red]; [.{}  \edge [red]; [.{} ]]]]]]]]]]]]]]]]]]]]]]] \edge [red]; [.{}  \edge [red]; [.{}  \edge [red]; [.{}  \edge [red]; [.{}  \edge [red]; [.{}  \edge [red]; [.{}  \edge [red]; [.{}  \edge [red]; [.{}  \edge [red]; [.{}  \edge [red]; [.{}  \edge [red]; [.{}  \edge [red]; [.{}  \edge [red]; [.{}  \edge [red]; [.{}  \edge [red]; [.{}  \edge [red]; [.{}  \edge [red]; [.{}  \edge [red]; [.{}  \edge [red]; [.{}  \edge [red]; [.{}  \edge [red]; [.{}  \edge [red]; [.{}  \edge [red]; [.{} ]]]]]]]]]]]]]]]]]]]]]] \edge [red]; [.{}  \edge [red]; [.{}  \edge [red]; [.{}  \edge [red]; [.{}  \edge [red]; [.{}  \edge [red]; [.{}  \edge [red]; [.{}  \edge [red]; [.{}  \edge [red]; [.{}  \edge [red]; [.{}  \edge [red]; [.{}  \edge [red]; [.{}  \edge [red]; [.{}  \edge [red]; [.{}  \edge [red]; [.{}  \edge [red]; [.{}  \edge [red]; [.{}  \edge [red]; [.{}  \edge [red]; [.{}  \edge [red]; [.{}  \edge [red]; [.{}  \edge [red]; [.{} ]]]]]]]]]]]]]]]]]]]]]]]]]\end{tikzpicture}} 
\end{figure}
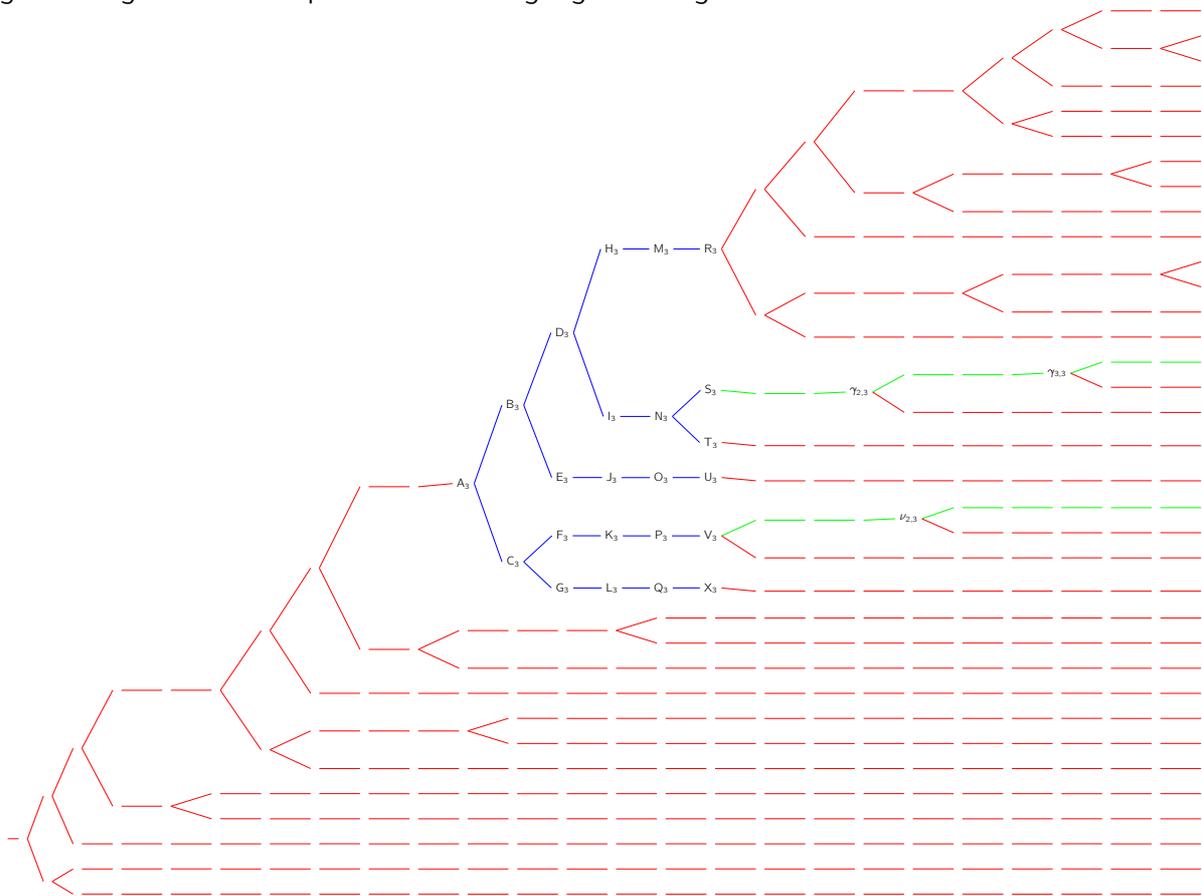

In Figure~\ref{fig:example-m6-tau3-g31}
we draw the edges and nodes of $\upsilon$ corresponding to semigroups  from genus $5$ up to genus $29$. The tree $\tau_3$ and the infinite chains $\zeta_3$ and $\eta_3$ are highlighted. Notice also that one can observe five full self-replications of the tree from Figure~\ref{fig:tree-m6}. 
The figure was generated with the {\tt drawsgtree} tool \cite{drawsgtree} with the command \textsc{./drawsgtree -g29 -m6 -e infinitechains -etrim -d .1}.

The number $i_g(6)$ of numerical semigroups of genus $g$ and multiplicity $6$ that are in an infinite chain whenever $g\geq5$, is equal to the number of numerical semigroups on level $g-5$ of the tree $\upsilon$. From what has been developed in Subsection~\ref{subsec-m6}, the growth of the tree $\upsilon$ is known and we are ready to state it in the next theorem.

  \begin{lemma}\label{lemma:msix}
The number $i_g(6)$ of numerical semigroups of genus $g\geq 5$ and multiplicity $6$ that are in an infinite chain is 
$$\left\{\begin{array}{l}
g-5-\left\lfloor\dfrac{g-5}{5}\right\rfloor
+\sum\limits_{ n=1}^{\left\lfloor\frac{g-2}{9}\right\rfloor} n+
    \sum\limits_{ n=\left\lfloor\frac{g+7}{9}\right\rfloor}^{\left\lfloor\frac{g-6}{5}\right\rfloor}\left\lfloor\dfrac{g-2-5 n}{4}\right\rfloor+\sum\limits_{n=2}^{\left\lfloor\frac{g+1}{9}\right\rfloor} (n-1)+
    \sum\limits_{ n=\left\lfloor\frac{g+10}{9}\right\rfloor}^{\left\lfloor\frac{g-7}{5}\right\rfloor}\left\lfloor\dfrac{g-3-5n}{4}\right\rfloor \\ 
\mbox{if }g\equiv1,2\mod 5,\\
g-4-\left\lfloor\dfrac{g-5}{5}\right\rfloor
+\sum\limits_{ n=1}^{\left\lfloor\frac{g-2}{9}\right\rfloor} n+
    \sum\limits_{ n=\left\lfloor\frac{g+7}{9}\right\rfloor}^{\left\lfloor\frac{g-6}{5}\right\rfloor}\left\lfloor\dfrac{g-2-5 n}{4}\right\rfloor+\sum\limits_{ n=2}^{\left\lfloor\frac{g+1}{9}\right\rfloor} (n-1)+
    \sum\limits_{ n=\left\lfloor\frac{g+10}{9}\right\rfloor}^{\left\lfloor\frac{g-7}{5}\right\rfloor}\left\lfloor\dfrac{g-3-5n}{4}\right\rfloor \\
\mbox{if }g\equiv0,3,4\mod 5.
\end{array}\right.
$$

\end{lemma}

\begin{proof}
  First of all, notice that the number of self-replications of the structure of $\tau_1$, from Figure~\ref{fig:tree-m6}, which lie before level $g-5$ of $\upsilon$ is $\lfloor\frac{g-5}{5}\rfloor$. 

For fixed $g$ we can define four sets of numerical semigroups:
\begin{align*}
  A&:=\left\{\Lambda:\Lambda \in \tau_{\left\lfloor\frac{g-5}{5}\right\rfloor+1} \text{ and } g(\Lambda)=g\right\} \\
B&:= \left\{\Lambda:\Lambda \text{ is a descendant of } K_n \text{ or } M_n, 1\leq n\leq \left\lfloor\frac{g-5}{5}\right\rfloor, \text{ and } g(\Lambda)=g\right\} \ \cup \\
& 
\left\{\Lambda:\Lambda \text{ is a descendant of } J_n \text{ or } L_n, 1\leq n\leq \left\lfloor\frac{g-5}{5}\right\rfloor, \Lambda\in \zeta_n\cup\eta_n,  \text{ and } g(\Lambda)=g\right\} \\
C&:= \left\{\Lambda:\Lambda \text{ is a descendant of } J_n, 1\leq n\leq \left\lfloor\frac{g-5}{5}\right\rfloor, \Lambda\notin \zeta_n,  \text{ and } g(\Lambda)=g\right\}\\
D&:= \left\{\Lambda:\Lambda \text{ is a descendant of } L_n, 2\leq n\leq \left\lfloor\frac{g-5}{5}\right\rfloor, \Lambda\notin \eta_n,  \text{ and } g(\Lambda)=g\right\}. 
\end{align*}

In the definition of $D$ the case $n=1$ has been excluded since $L_1$ belongs only to one infinite chain which is exactly $\eta_1$.

Now, the numerical semigroups of genus $g\leq 5$ and multiplicity $6$ which are in an infinite chain are the numerical semigroups of $A\sqcup B\sqcup C\sqcup D$.

To get the cardinality of $A$, consider the remainder $r$ such that $g-5=5\lfloor\frac{g-5}{5}\rfloor+r$. It is the level of the tree $\tau_{\left\lfloor\frac{g-5}{5}\right\rfloor+1}$ at which the numerical semigroups of genus $g$ that participate in self-replication lie. So, the cardinality of $A$ is equal to $r$ if $r=1,2$ and equal to $r+1$ when $r=0,3,4$.

By Proposition~\ref{prop:descandants-Tn-Xn}, Lemma~\ref{lemma:gamma}, and Lemma~\ref{lemma:nu}, the cardinality of $B$ is equal to $4\lfloor\frac{g-5}{5}\rfloor$.

On the other hand, $5\left\lfloor\frac{g-5}{5}\right\rfloor=g-5-r$. So,
$$\#A+\#B=\left\{\begin{array}{ll}
g-5-\lfloor\frac{g-5}{5}\rfloor&\mbox{if }g\equiv 1,2 \mod 5\\
g-4-\lfloor\frac{g-5}{5}\rfloor&\mbox{if }g\equiv 0,3,4 \mod 5.\\
\end{array}\right.$$

The descendants of $J_n$, for $1\leq n\leq \left\lfloor\frac{g-5}{5}\right\rfloor$, which belong to infinite chains but do not belong to $\zeta_n$ and have genus $g$, $g\geq 10$, are in bijection with the numerical semigroups $\gamma_{n,t}$ from Lemma~\ref{lemma:gamma} such that $g(\gamma_{n,t})<g$. Since $g(\gamma_{n,t})=5n+4t+1$, we have that the cardinality of $C$ is equal to 
  \begin{align*}
    & \#\left\{(n,t):5n+4t+1<g, \text{ where }1\leq n\leq \left\lfloor\frac{g-5}{5}\right\rfloor \text{ and } 1\leq t \leq n\right\} \\
    & \#\left\{(n,t):5n+4t+2\leq g, \text{ where }1\leq n\leq \left\lfloor\frac{g-5}{5}\right\rfloor \text{ and } 1\leq t \leq n\right\} \\
    &= \sum\limits_{ n=1}^{\left\lfloor\frac{g-2}{9}\right\rfloor} n+
    \sum\limits_{ n=\left\lfloor\frac{g+7}{9}\right\rfloor}^{\left\lfloor\frac{g-6}{5}\right\rfloor}\left\lfloor\frac{g-2-5n}{4}\right\rfloor. 
\end{align*}

On the other hand, by Lemma~\ref{lemma:nu}, the set $D$ is in bijection with the set $\{\nu_{n,t}:2\leq n  \leq \left\lfloor\frac{g-5}{5}\right\rfloor,1\leq t \leq n-1 \text{ and } g(\nu_{n,t})<g\}$. Furthermore, since $g(\nu_{n,t})=5n+4t+2$, we have that the cardinality of $D$ is equal to 
\begin{align*}
& \#\left\{(n,t):5n+4t+2<g, \text{ where }2\leq n\leq \left\lfloor\frac{g-5}{5}\right\rfloor \text{ and } 1\leq t \leq n-1\right\} \\
    & \#\left\{(n,t):5n+4t+3\leq g, \text{ where }2\leq n\leq \left\lfloor\frac{g-5}{5}\right\rfloor \text{ and } 1\leq t \leq n-1\right\} \\
    &= \sum\limits_{ n=2}^{\left\lfloor\frac{g+1}{9}\right\rfloor} (n-1)+
    \sum\limits_{ n=\left\lfloor\frac{g+10}{9}\right\rfloor}^{\left\lfloor\frac{g-7}{5}\right\rfloor}\left\lfloor\frac{g-3-5n}{4}\right\rfloor\\
    &=\sum\limits_{ n=1}^{\left\lfloor\frac{g-8}{9}\right\rfloor}n+
    \sum\limits_{ n=\left\lfloor\frac{g+10}{9}\right\rfloor}^{\left\lfloor\frac{g-7}{5}\right\rfloor}\left\lfloor\frac{g-3-5n}{4}\right\rfloor. 
\end{align*}
\end{proof}

The anonymous reviewer noticed that the formula given in Lemma~\ref{lemma:msix} is nothing else but $i_g(6)=\left\lfloor\frac{2g^2+g-16}{9}\right\rfloor-\left\lfloor\frac{(g-1)^2}{5}\right\rfloor.$
It can be proved by some arguments on quasi-polynomials or using the implementation of Barvinok's algorithm in LattE \cite{barvinok}.
Hence, we can state now the following theorem.
  
  \begin{theorem}\label{theorem:msix}
The number $i_g(6)$ of numerical semigroups of genus $g\geq 5$ and multiplicity $6$ that are in an infinite chain is $$i_g(6)=\left\lfloor\frac{2g^2+g-16}{9}\right\rfloor-\left\lfloor\frac{(g-1)^2}{5}\right\rfloor.$$
    \end{theorem}

\section{Open question}
  
Proposition 7 in \cite{kaplan} states that the number of all semigroups of genus $g$ with fixed multiplicity $m$ is eventually a quasi-polynomial in $g$. The results presented in this work suggest, as pointed out by the anonymous referee, the following conjecture.

\begin{conjecture}
  For a fixed multiplicity $m$, the number of semigroups of genus $g$ lying in an infinite chain is quasi-polynomial in $g$, for $g\geq m-1$.
\end{conjecture}

\section{Acknowledgments}
The authors want to express their great gratitude to the referee of the manuscript. They specially appreciate the simplification of the expression of $i_6(g)$, as well as suggesting the conjecture at the end of the work.

The first author was supported by the Catalan government under grant 2021 FISDUR 00189. Both authors were supported by the Spanish government under grant PID2021-124928NB-I00, and by the Catalan government under grant 2021 SGR 00115.

All the graphs have been drawn using the open {\tt drawsgtree} tool, which can be downloaded from {\tt https://github.com/mbrasamoros/drawsgtree} (\cite{drawsgtree}).

\end{document}